# Anomalous Hall effect arising from noncollinear antiferromagnetism

Hua Chen[1]*, Qian Niu[1], A. H. MacDonald[1]

[1]Department of Physics, University of Texas at Austin, Austin, TX 78712, USA.

*Email: huachen@physics.utexas.edu.

In most conductors current flow perpendicular to electric field direction (Hall current) can be explained in terms of the Lorentz forces ($\frac{e}{c}\vec{v}\times\vec{B}$) present when charged particles flow in an external magnetic field. However, as established in the very early work of Edwin Hall, ferromagnetic conductors such as Fe, Co, and Ni have an anomalous Hall conductivity contribution that cannot be attributed to Lorentz forces and therefore survives in the absence of a magnetic field[1,2]. Although the anomalous Hall effect is experimentally strong, it has stood alone among metallic transport effects for much of the last century because it lacked a usefully predictive, generally accepted theory. Progress over the past decade has explained why. It is now clear that the anomalous Hall effect in ferromagnets has contributions[2] from both extrinsic scattering mechanisms similar to those that determine most transport coefficients, and from an intrinsic mechanism that is independent of scattering[3-5]. The anomalous Hall effect is also observed in paramagnets, which have nonzero magnetization induced by an external magnetic field. Although no explicit relationship has been established, the anomalous Hall effect in a particular material is usually assumed to be proportional to its magnetization. In this work we point out that it is possible to have an anomalous Hall effect in a noncollinear antiferromagnet with zero net magnetization provided that certain common symmetries are absent, and predict that $Mn_3Ir$, a technologically important antiferromagnetic material with noncollinear order that survives to very high temperatures, has a surprisingly large anomalous Hall effect comparable in size to those of the elemental transition metal ferromagnets.

$Mn_3Ir$ can be viewed as an fcc crystal with Mn atoms on three of the four cubic sublattices (Fig. 1a), and is widely used as an exchange-bias material in spin-valve devices. Its Mn sublattice can be viewed as consisting of two-dimensional kagome lattices (Fig. 1b) stacked along the (111) direction. Although isolated two-dimensional kagomes order weakly because of strong magnetic frustration[6-10], the Mn moments in $Mn_3Ir$ have strong three-sublattice triangular (T1) magnetic order[11] (Fig. 1a) and a high magnetic transition temperature ~950 K. The surprising stability is provided by inter-kagome coupling[12] combined with strong magnetic anisotropy[13]. Because the kagome structure, with appropriate exchange interactions, can support relatively simple noncollinear antiferromagnetic states similar to those of $Mn_3Ir$, we begin our discussion of the anomalous Hall effect in noncollinear antiferromagnets by focusing on a simple two-dimensional kagome model.

The itinerant electron Hamiltonian for an *s-d* model on the kagome lattice is

$$H = t \sum_{<i\alpha, j\beta>\gamma} c^+_{i\alpha\gamma} c_{j\beta\gamma} - J \sum_{i\alpha\gamma\tau} c^+_{i\alpha\gamma} c_{i\alpha\tau} \vec{\sigma}_{\gamma\tau} \cdot \vec{\Delta}_\alpha . \qquad (1)$$

Here $\alpha$ and $\beta$ label kagome sublattices, the angle-brackets restrict *s*-electron hopping to nearest neighbor sites, and $\vec{\Delta}_\alpha$ is an exchange field due to coupling to quenched *d*-electron local moments that have coplanar triangular magnetic order (Fig. 1b). The band structure of this model is shown in Fig. 2a. One interesting feature is that, in contrast to the case of simple collinear bipartite antiferromagnets[14], the bands are not doubly degenerate. In the latter case, the time-reversal operation *T* in combination with a spatial translation by the vector connecting the two sublattices $T_{\vec{d}}$ is a good symmetry, supplying an analog of Kramer's theorem even though time-reversal symmetry on its own is broken. When combined with spatial inversion symmetry *I*, this property implies double band degeneracy at every point in momentum space[14]. In contrast, in the three-sublattice kagome model there is no operation *O* that can make *TO* a good symmetry while fulfilling the prerequisites for Kramer's theorem, i.e., that *TO* is antiunitary and $(TO)^2 = -1$.

Mirror reflection *M* with respect to the kagome plane is a good symmetry of the kagome lattice, but flips all in-plane spin components. *M* symmetry is therefore broken by the term describing the interaction with local moments $\vec{\Delta}_\alpha$ in equation (1). Since time reversal *T* reverses the in-plane spins a second time, the Hamiltonian in equation (1) does have *TM* symmetry. However, this symmetry does not imply double band degeneracy because $(TM)^2 = 1$ rather than -1.

The fact that *TM* is a good symmetry of our model does have other important consequences as we now show. Momentum space Berry curvature can be used to characterize nontrivial topological properties of condensed matter systems and is directly related to the anomalous Hall effect[4,15]. The momentum space Berry curvature of a Bloch band is defined as $\vec{\Omega}_n(\vec{k}) = i\langle \nabla_{\vec{k}} u_{n\vec{k}} | \times | \nabla_{\vec{k}} u_{n\vec{k}} \rangle$, where $|u_{n\vec{k}}\rangle$ is the Bloch eigenstate for band *n* at momentum $\vec{k}$. The semiclassical velocity formula for charge carriers is modified when Berry curvature is nonzero:

$$\dot{\vec{r}}_n(k) = \frac{\partial \varepsilon_{n\vec{k}}}{\hbar \partial \vec{k}} - \dot{\vec{k}} \times \vec{\Omega}_n(\vec{k}). \quad (2)$$

Since $\dot{\vec{k}}$ is proportional to electric field, the intrinsic anomalous Hall effect is proportional to the Brillouin-zone integration of the Berry curvature[2,15]. If a system has both time reversal *T* and spatial inversion *I* symmetry, the Berry curvature satisfies $\vec{\Omega}_n(\vec{k}) = -\vec{\Omega}_n(-\vec{k})$ and $\vec{\Omega}_n(\vec{k}) = \vec{\Omega}_n(-\vec{k})$, implying that it vanishes identically. In the present case, *TM* symmetry implies that $\Omega_n^z(\vec{k}) = -\Omega_n^z(-\vec{k})$, and *I* symmetry that $\Omega_n^z(\vec{k}) = \Omega_n^z(-\vec{k})$, so the Berry curvature still vanishes identically. An explicit evaluation of $\Omega_n^z(\vec{k})$ for the model of equation (1) confirms this conclusion, except at points in the Brillouin zone where degeneracies occur and $\Omega_n^z(\vec{k})$ is not well defined (Fig. 2a). The presence of the band-crossing (Dirac) points in Fig. 2a, however, suggests that the system has the potential to become topologically nontrivial when *TM* symmetry is broken.

We can conclude from the discussion above that the Berry curvature $\Omega_n^z(\vec{k})$ will be an even nonzero function of $\vec{k}$ if we break *TM* symmetry, and that the intrinsic anomalous Hall conductivity will therefore be non-zero. A similar symmetry argument applies to the extrinsic anomalous Hall effect as well. One way to break *TM* symmetry is to add an external magnetic field $\vec{B}$ perpendicular to the plane, which breaks *T* symmetry but, since $\vec{\sigma}$ is a pseudovector, not *M* symmetry. The influence of a magnetic field can therefore be captured by adding a Zeeman term to equation (1) (We ignore any change in the in-plane components of $\vec{\Delta}_\alpha$):

$$H_B = -J_z \sum_{i\alpha\gamma\tau} c^+_{i\alpha\gamma} c_{i\alpha\tau} \sigma^z_{\gamma\tau}. \qquad (3)$$

As can be seen from the band structure in Fig. 2b, the Dirac-point degeneracies are lifted by this perturbation and the Berry curvature (Fig. 2c) becomes nonzero. In contrast to the case of ferromagnetic metals, this anomalous Hall effect mechanism does not require spin-orbit coupling, and is in this sense reminiscent of the real-space *spin chirality* topological Hall effect[16,2] scenario. However, it is *TM* symmetry breaking by the Zeeman term equation (3) rather than the noncoplanar configuration of the background moments that is responsible for the anomalous Hall effect. We also note that a nonzero total magnetization is also induced by the Zeeman term.

We now show that broken mirror symmetry combined with spin-orbit coupling leads to an anomalous Hall effect, but not necessarily to a net magnetization. Spin-orbit coupling is *T* invariant, but can break *M* symmetry and therefore the combined *TM* symmetry on a kagome lattice. For example, consider the following spin-orbit coupling term:

$$H_{SO} = \iota t_{SO} \sum_{<i\alpha,j\beta>\gamma\tau} v_{\alpha\beta} \vec{\sigma}_{\gamma\tau} \cdot \vec{n}_{\alpha\beta} c^+_{i\alpha\gamma} c_{j\beta\tau},  \qquad (4)$$

where $t_{SO}$ is related to the atomic spin-orbit coupling, $\iota = \sqrt{-1}$, $v_{\alpha\beta}$ is antisymmetric with respect to its sublattice indices and $v_{12} = v_{23} = v_{31} = 1$, and $\vec{n}_{\alpha\beta}$ are a set of coplanar unit vectors shown in Fig. 2d. The spin-orbit coupling term identified in equation (4) is analogous to the corresponding term in the graphene model of Ref. [17] and preserves the inversion symmetry, but breaks the mirror symmetry. It accounts for the difference between the left-hand and right-hand environments of an electron hopping between nearest neighbors on a kagome lattice. When $H_{SO}$ is added to the model of equation (1), gaps open at the Dirac points (Fig. 2e), as in the external magnetic field case (Fig. 2b). Figure 2f shows that the Berry curvature is again non-zero throughout momentum space, and that it is large near the gapped Dirac points.

This calculation proves by explicit example that antiferromagnets can have non-zero anomalous Hall conductivities simply because they break time reversal symmetry. This property has largely[18-20] escaped notice because most antiferromagnets have other symmetries that force their Hall conductivities to vanish. For example consider the case of collinear antiferromagnets on bipartite lattices with inversion *I* symmetry discussed previously in connection with band spin-splitting. In the absence of an external magnetic field, $TT_{\vec{d}}$ is a good symmetry even with spin-orbit coupling, and the Berry curvature and Hall conductivity vanish. When a small external magnetic field is applied and spin-orbit interactions are negligible, the antiparallel moments on opposite sublattices rotate slightly from the perpendicular-to-field plane toward the field direction. The system is still invariant under the combined operation of $T$, $T_{\bar{d}}$ and a collective rotation of all spins by $\pi$ around the perpendicular-plane spin-projections: $R_{\bar{n}}^{\pi}$. Therefore even though spin-degeneracy is lifted by the external magnetic field, $TT_{\bar{d}}R_{\bar{n}}^{\pi}$ symmetry plus $I$ symmetry still guarantees vanishing Berry curvature. Once spin-orbit coupling is added, however, $R_{\bar{n}}^{\pi}$ will inevitably influence the orbital wave functions so that $TT_{\bar{d}}R_{\bar{n}}^{\pi}$ is no longer a good symmetry. We conclude that when an external magnetic field is present antiferromagnets in this class will have an anomalous Hall effect similar to those induced in a paramagnetic system by spin-orbit coupling. On the other hand, the operations that might leave the system invariant when combined with $T$ are more limited in the case of noncollinear antiferromagnets.

What are the implications of these considerations for $Mn_3Ir$, which has inversion symmetry, but lacks (111) plane mirror symmetry because of the fcc stacking of its kagome layers (Fig. 1a)? Ir is a heavy element whose large atomic spin-orbit coupling will be transferred to the Mn atoms with which it hybridizes. Our symmetry considerations therefore suggest that antiferromagnetic $Mn_3Ir$ should have a sizeable anomalous Hall effect in the absence of a magnetic field. To explore this possibility we have used first-principles methods to evaluate its intrinsic anomalous Hall conductivity. (Technical details of this calculation are explained in Methods and Supplementary Information).

First of all we found that fully-relativistic electronic structure calculations predict that the three Mn moments in the $Mn_3Ir$ unit cell form a very small angle $\approx 0.1°$ relative to the (111) plane, leading to a total magnetization of ~0.02 $\mu_B$ per formula unit along the (111) direction. This moment rotation can be traced to magneto-crystalline anisotropy effects[13]. From a symmetry point of view, these small rotations, like the anomalous Hall effect discussed below, are a combined consequence of mirror symmetry breaking and spin-orbit coupling. By performing the Brillouin-zone integral of the (111) direction Berry curvature component, we find that the corresponding intrinsic anomalous Hall conductivity is $\sigma_{(111)}$=218 $\Omega^{-1}$cm$^{-1}$, the same order of magnitude as for the ferromagnetic transition metals Fe, Co, Ni. As in the ferromagnetic metal case[21], the band sum of the Berry curvature is strongly peaked at points in momentum space where degeneracies (Fig. 3a) are lifted by spin-orbit coupling and states are split to opposite sides of the Fermi energy (Fig. 3b). To show that the anomalous Hall conductivity is not due to the small net magnetization, we have constrained the Mn moments to be strictly coplanar, and found that $\sigma_{(111)}$ changes by only 1 $\Omega^{-1}$cm$^{-1}$. When spin-orbit coupling is not included, the anomalous Hall conductivity vanishes. These results are consistent with the symmetry arguments explained above using the kagome lattice example, and show that the symmetry allowed anomalous Hall effect in Mn$_3$Ir is surprisingly large.

In the absence of spin-orbit coupling the broken symmetry ground state of Mn$_3$Ir is invariant under any global spin rotation. With spin-orbit coupling, the ground state has a discrete two-fold degeneracy. The magnetic configuration sketched in Fig. 1a, which we call A phase from now on, is degenerate with its time-reversal counterpart, which we call B phase, and with other phases associated with other (111)-type planes. The small net magnetizations and the large anomalous Hall conductivities of the A and B phases are opposite. If one adds an external magnetic field along the (111) direction, the nearly coplanar Mn moments will rotate toward the field, which changes the anomalous Hall conductivity only slightly as discussed above.

To develop a comprehensive picture of all these effects, we constrained the Mn moments of both A and B phases to have different angles relative to the (111) plane. Figure 3c shows the dependence of the total energy (relative to the ground states) of the two phases on the rotation angle. The large energy changes, by hundreds of meV per formula unit, reflect the strong exchange coupling between Mn moments[13], and imply that only small rotation angles can be achieved using magnetic fields. The energy differences between A and B phases at a given rotation angle (Fig. 3d) reflect the absence of mirror symmetry. Figure 3e plots intrinsic anomalous Hall conductivities calculated for various magnetic configurations. The detailed shape of these curves is sensitively dependent on band structure (Supplementary Figs. S2 and S3); large changes in $\sigma_{(111)}$ signal that the Fermi energy has entered or exited the energy interval between spin-orbit split states. Note that the anomalous Hall effect is due almost entirely to spin-orbit coupling even when the moments are tilted out of (111) planes.

To experimentally observe the anomalous Hall effect in $Mn_3Ir$, it will be necessary to prepare a sample in which one magnetic phase is dominant. Since all phases have small but non-zero magnetizations, one strategy would be to apply an external magnetic field oriented along a (111) direction at the highest convenient temperature and then cool. Another strategy would be to study instead the closely related material $Mn_3Pt$, which has collinear and noncollinear antiferromagnetic phases in different temperature ranges[22]. Crossing a collinear to non-collinear phase boundary under the influence of a magnetic field could also selects a dominant noncollinear phase. Accurate stoichiometry may also be important; the $L1_0$ phase of MnIr, for example, is a collinear antiferromagnet that should not have an anomalous Hall effect. If single-phase materials could be prepared, their magnetization configuration could be read electrically, or, since the anomalous Hall effect is always accompanied by optical gyrotropy[23], optically.

**Methods:** First-principles electronic structure calculations for $Mn_3Ir$ were carried out using the QUANTUM ESPRESSO package[24]. Fully relativistic norm-conserving pseudopotentials under the Perdew, Burke, and Ernzerhof generalized-gradient approximation[25] for Mn and Ir are generated with the ATOMIC code included in the QUANTUM ESPRESSO package. The lattice constant of $Mn_3Ir$ is set to the experimental value of 3.785 Å [26]. A kinetic-energy cutoff of 180 Ry (720 Ry for the charge densities) is used for plane waves. The ground state is obtained from a self-consistent calculation using a 12×12×12 Monkhorst-Pack $k$-point mesh[27] and a convergence threshold of $10^{-8}$ Ry. We found in the ground state each Mn atom has a magnetic moment of 2.91 $\mu_B$, slightly larger than previous first-principles results[13,28], and the Ir atom has vanishing magnetic moment. To simulate the effect of an external magnetic field along the (111) direction, a penalty potential that forces the magnetic moment of each atom to be at desired directions is added in the self-consistent calculation. The anomalous Hall conductivity is calculated utilizing the Wannier interpolation approach[29-31]. Further details of the calculations are available in the Supplementary Information.

**Acknowledgments:** H.C. thanks Lifa Zhang, Xiao Li, and Karin Everschor for valuable discussions. H.C. and A.H.M. were supported by the Welch Foundation under Grant No. TBF1473 and by the National Science Foundation under Grant No. DMR-1122603. Q.N. was supported by the US Department of Energy Division of Materials Sciences and Engineering under Grant No. DE-FG03-02ER45958. The calculations were mainly performed at Texas Advanced Computing Center.

**Figure Legends:**

**Figure 1 | Structure of Mn$_3$Ir. a**, Unit cell of Mn$_3$Ir with triangular antiferromagnetic order. **b**, An individual (111) plane of Mn$_3$Ir. The Mn atoms form a kagome lattice.

**Figure 2 | Noncollinear antiferromagnet on a kagome lattice. a**, Band structure of the model in equation (1), $J=1.7t$. **b**, Band structure with magnetic field, $J_z=0.2t$. **c**, Berry curvature of the lowest band in **b**. **d**, Vectors $\vec{n}_{\alpha\beta}$ for the spin-orbit coupling term in equation (4). **e**, Band structure with spin-orbit coupling, $t_{SO}=0.2t$. **f**, Berry curvature of the lowest band in **e**.

**Figure 3 | Anomalous Hall effect of Mn$_3$Ir. a**, Band structure of the A phase near the Fermi level. **b**, Berry curvature along the same *k*-point path as **a**. **c**, Relative energies of the A and B phases with Mn moment angle relative to the (111) plane constrained. The ground states at ±0.1° are shifted horizontally for illustration purpose. **d**, Energy differences between A and B phases at different angles. **e**, Anomalous Hall conductivities of A and B phases (black squares) with and without spin-orbit coupling (red triangles) as a function of the Mn moment tilt angle.

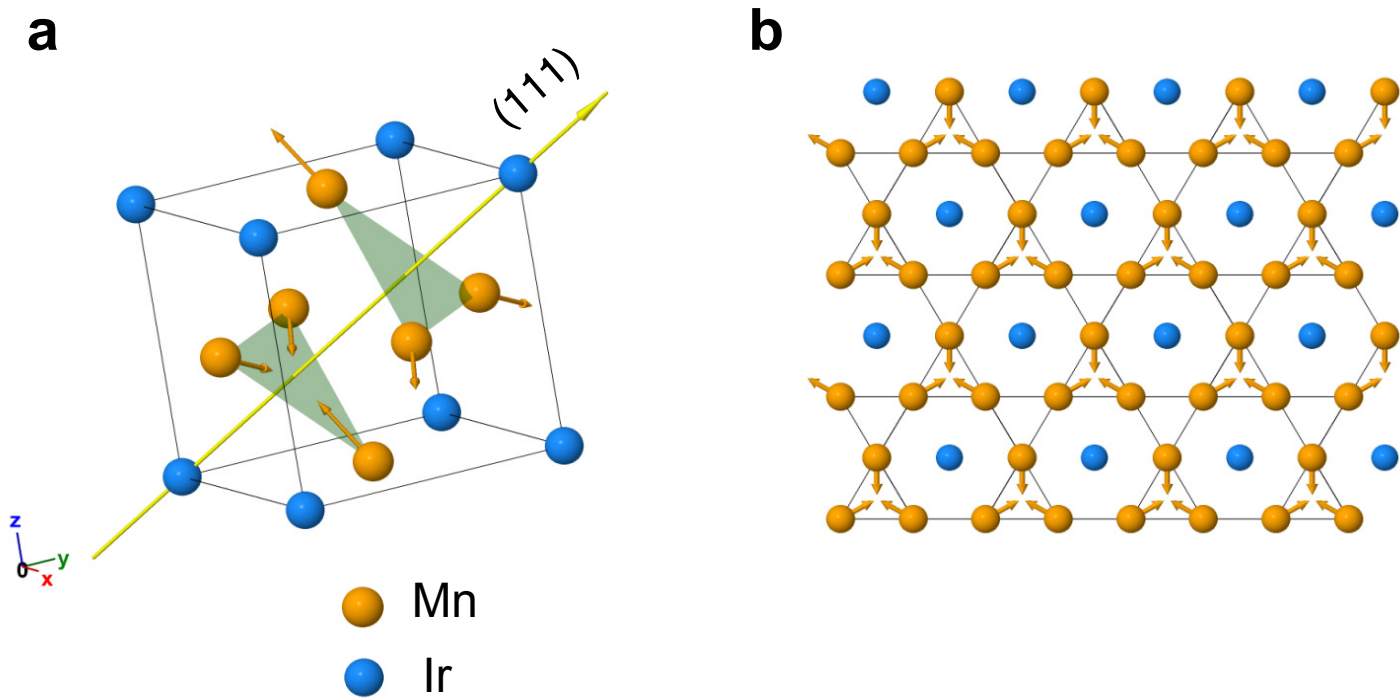

Figure 1. Chen et al.

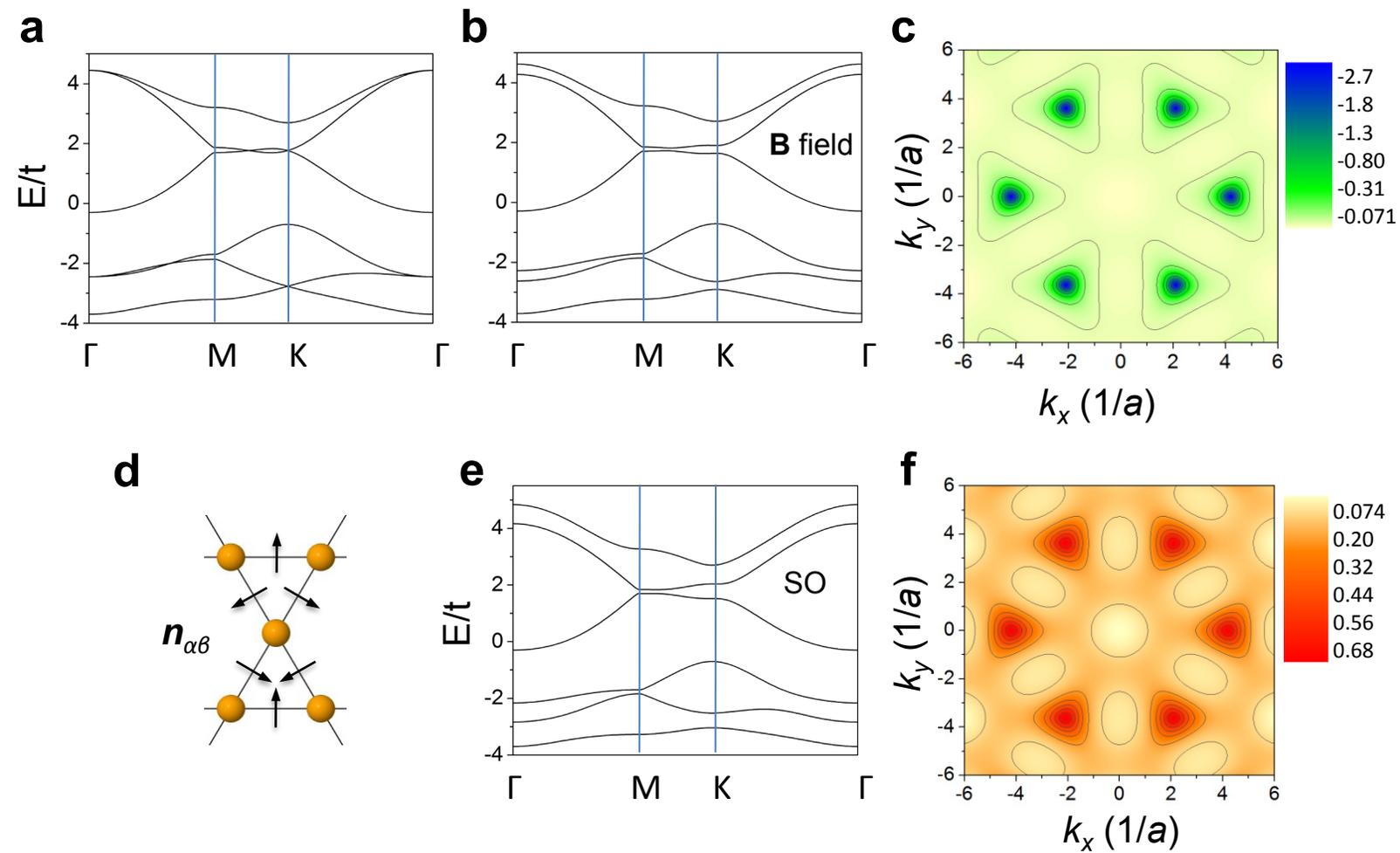

Figure 2. Chen et al.

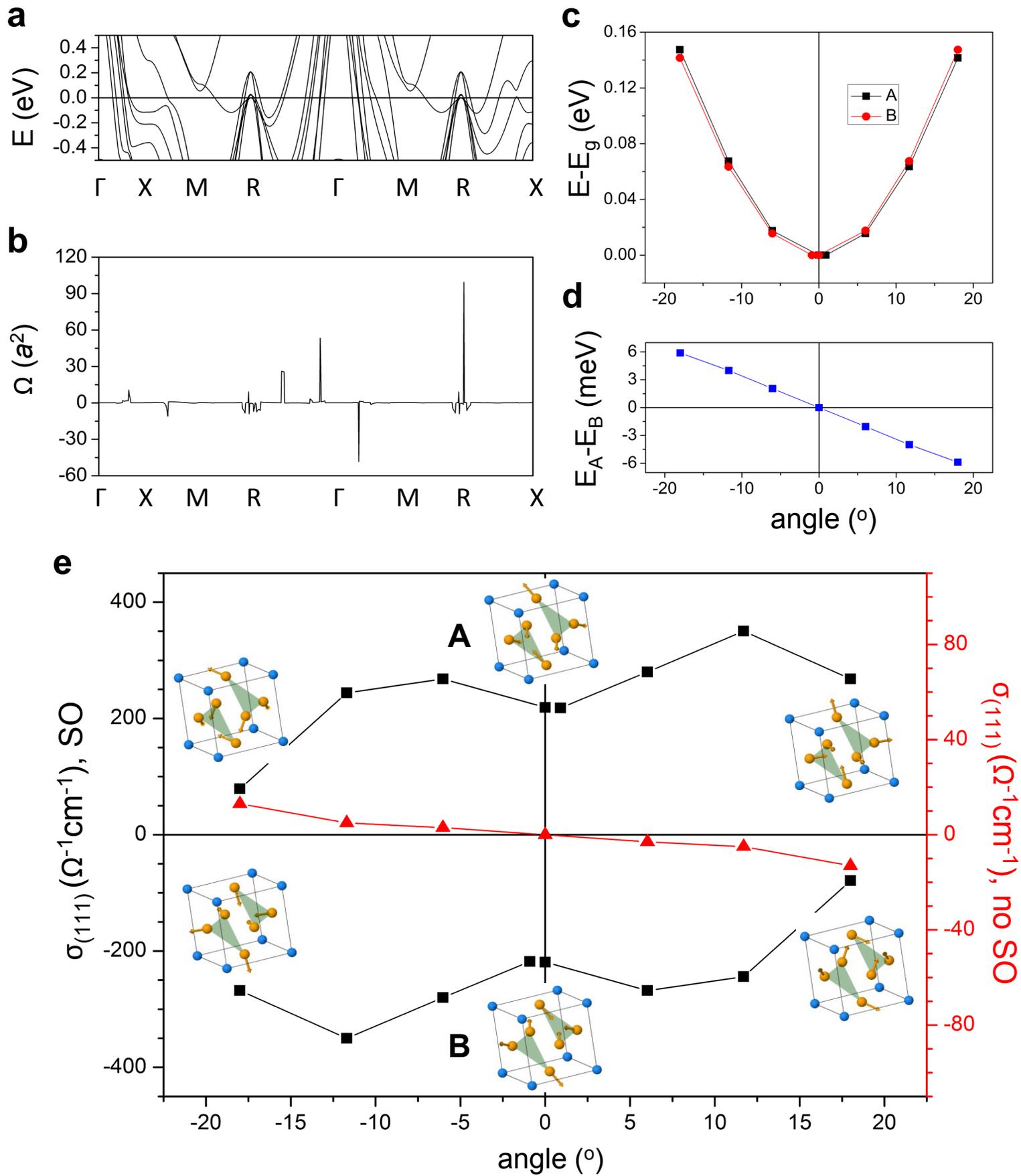

Figure 3. Chen et al.